\documentclass[%
preprint, showpacs,preprintnumbers,
nofootinbib, nobibnotes,
 amsmath,amssymb,
 aps,
]{revtex4-1}

\usepackage{graphicx}
\usepackage{dcolumn}
\usepackage{bm}

\begin{document}

%\preprint{APS/123-QED}

\title{ Manifestation of nuclear cluster structure in Coulomb sums }

\author{A.Yu.~Buki}
 \email{abuki@ukr.net}
\affiliation{%
 National Science Center "Kharkov Institute of
Physics and Technology" \\ 61108, Kharkov, Ukraine\\}%
\author{I.S.~Timchenko}
 \email{timchenko@kipt.kharkov.ua}
 \affiliation{%
 National Science Center "Kharkov Institute of
Physics and Technology" \\ 61108, Kharkov, Ukraine\\}%

\date{\today}

\begin{abstract}
Experimental Coulomb sum values of $^6$Li and $^7$Li nuclei have
been obtained, extending the earlier reported momentum transfer
range of Coulomb sums for these nuclei up to $q = 0.750 \div 1.625 $
~fm$^{-1}$. The dependence of the Coulomb sums on the momentum
transfers of $^6$Li and $^7$Li is shown to differ substantially from
similar dependences for all the other  nuclei investigated.
Relationship between the nuclear cluster structure and Coulomb sums
has been considered. The momentum transfer value, above which the
Coulomb sum becomes constant, is found to be related to the cluster
isolation parameter $x$, which characterizes the degree of nuclear
clusterization.
\begin{description}

\item[PACS numbers]
21.60.Gx, 25.30.Fj, 27.20.+n
\item[KEY WORDS] 
Coulomb sum, lithium isotopes, light and medium nuclei, 
nuclear cluster structure, cluster isolation parameter. 
\end{description}
\end{abstract}

%\pacs{Valid PACS appear here}

\maketitle

\section{ Introduction }\label{introd}
In the double-differential cross-section for electron scattering by
the nucleus, \mbox{$\left(d^2{\rm \sigma}/d{\rm \Omega}d\rm
\omega\right)$}, the contributions from the electron-nucleus
interaction may be separated by means of longitudinal and transverse
components of the electromagnetic field. Accordingly, these
contributions are called the longitudinal and transverse response
functions ($R_{\rm L}(q,\omega)$ and $R_{\rm T}(q,\omega)$,
respectively). According to ref.~\cite{1} , the double-differential
cross-section is related to the response functions by the equation
% \begin{equation}
% \frac{d^2\sigma}{d\Omega d\omega }(\theta,E_0,\omega) =
%\sigma_{\rm M}(\theta,E_0)\!\left[\frac{Q^4}{q^4}R_L(q,\omega)+
%\left(\frac{Q^2}{2q^2}+\tan^2\frac{\theta}{2}\right)R_T(q,\omega)\right].\label{eq1}
 %\end{equation}
 \begin{eqnarray}
 \frac{d^2\sigma}{d\Omega d\omega }(\theta,E_0,\omega) =  \sigma_{\rm
  M}(\theta,E_0)\times  \nonumber \\
\!\left[\frac{Q^4}{q^4}R_{\rm L}(q,\omega)+
\left(\frac{Q^2}{2q^2}+\tan^2\frac{\theta}{2}\right)R_{\rm T}(q,\omega)\right],\label{eq1}
\end{eqnarray}
where  $\omega$, $q$, $Q = (q^2 -  \omega^2)^{1/2}$ are,
respectively, the energy, 3-momentum, 4-momentum transferred to the
nucleus by the incident electron of initial energy $E_0$ and
scattered by the angle $\theta$; \mbox{$\sigma_{\rm M}(\theta,E_0) = e^4\cos^2
(\theta/2) / [4 E_0^2 \sin^4(\theta/2)]$} is the Mott cross-section;
$e$ is the electron charge.

In the treatment of the experimental data, one must take into
account the influence of the nuclear electrostatic field on the
incident electron. For this purpose, the correction  $\Delta E_0$ is
introduced into the definition of the 3-momentum transfer $q =
\{4(E_0 +  \Delta E_0) [(E_0 + \Delta E_0) - \omega]
\sin^2(\theta/2) + \omega^2 \} ^{1/2}$. The correction $\Delta E_0$
is given by \mbox{$k(3/2)Ze^2/R$}, where $R$ is the radius of the
equivalent homogeneous distribution. According to ref.~\cite{2}, for
electrons scattered by light nuclei to the continuum region the
coefficient $k$ is equal to 0.8.

The experimental data on the longitudinal functions $R_{\rm L}(q,\omega)$
are generally represented as Coulomb sums
\begin{equation}
S_{\rm L}(q) =\frac{1}{Z}\int_{\omega_{\rm el}^+}^\infty
\frac{R_{\rm L}(q,\omega)}{\eta~[\tilde{G}_{\rm E} (Q^2)]^2} d\omega
, \label{eq2}
\end{equation}
where  \mbox{$\left[\tilde{G}_{\rm E}(Q^2)\right]^2 = \left[{G}_{\rm
E}^{\rm p}(Q^2)\right]^2+\frac{N}{Z}\left[{G}_{\rm E}^{\rm
n}(Q^2)\right]^2$}. Here, \mbox{$\omega_{\rm el}^+$}, being the
lower limit of integral~(\ref{eq2}), corresponds to the energy
transfer of the elastic electron scattering peak, and the
superscript "+" excludes the contribution of this peak to the
integral; $N$ and $Z$ denote the number of neutrons and protons in
the nucleus, respectively; \mbox{$\eta =
\left[1+Q^2/(4M^2)\right]\times\left[1+Q^2/(2M^2)\right]^{-1}$} is
the correction for the relativistic effect of nucleon motion in the
nucleus; $M$ is the proton mass; $G_{\rm E}^{\rm p}$ and  $G_{\rm
E}^{\rm n}$ are the charge form factors of the proton and the
neutron, respectively.

\begin{figure}[bh] %\noindent
\centering{
\includegraphics[width=90mm]{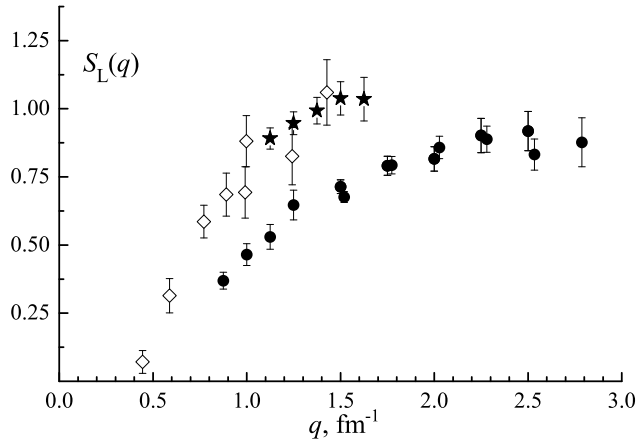}
} \caption{Coulomb sums of $^4$He and $^6$Li nuclei. Full circles -- $^4$He \cite{3, 4, 5}, 
diamonds -- $^6$Li \cite{6}, full asterisks -- $^6$Li \cite{7}.}
\label{fig1}
\end{figure}

For all the nuclei studied, the behavior of \mbox{$S_{\rm L}(q)$}
with variations in the momentum transfer is similar in its
character. With an increase in $q$, the function \mbox{$S_{\rm
L}(q)$} increases until at a certain momentum transfer value,
denoted as $q_{\rm p}$, the \mbox{$S_{\rm L}(q)$} takes on constant
values forming the function \mbox{$S_{\rm L}(q)$} plateau. For
almost all previously studied nuclei we have $q_{\rm p} \approx
2$~fm$^{-1}$. By way of illustration, Fig.~\ref{fig1} shows the
experimental \mbox{$S_{\rm L}(q)$} values for the $^4$He nucleus
\cite{3, 4, 5}.

The authors of papers \cite{6,7} have determined \mbox{$S_{\rm
L}(q)$} values for the $^6$Li nucleus, and have found that the
behavior of the function differs from the usual one
(Fig.~\ref{fig1}). It can be seen that the \mbox{$S_{\rm L}(q)$}
function reaches the plateau at $q_{\rm p} \approx 1.4$~fm$^{-1}$,
this being much earlier in $q$ than in the case with $^{4}$He and
other nuclei. In the $^7$Li case, in the measurement range $q =
1.250 \div 1.625$ fm$^{-1}$ (see ref.~\cite{8}), the function
\mbox{$S_{\rm L}(q)$} is constant within the experimental error. It
means that if the \mbox{$S_{\rm L}(q)$} value is lower at certain
momentum transfers, then it will reach the plateau range at $q_{\rm
p} \leq 1.3$ fm$^{-1}$. Thus, the data of ref.~\cite{8} do not
specify $q_{\rm p}$ for the $^7$Li nucleus, but restrict its upper
value. The authors of works \cite{7, 8} have put forward the
hypothesis that a comparatively low $q_{\rm p}$ value in the
$^{6,7}$Li case may be due to the Coulomb sum manifestation of
clusterization peculiar to the nuclei under discussion.

However, on a more rigorous approach to the problem of relationship
between the $q_{\rm p}$ value and nuclear cluster structure it
should be noted that this hypothesis is actually based only on the
experimental $q_{\rm p}$ value of the $^6$Li nucleus. As regards the
$q_{\rm p}$ value of $^7$Li, from the data of \cite{8} it follows
that it is not higher than that of $^6$Li, and it is not improbable
that it may be substantially lower. The last version would be in
contrast with the proposed hypothesis, because if the $q_{\rm p}$
value is related to the clusterization (and the nuclei $^6$Li and
$^7$Li are close in the degree of clusterization), then the $q_{\rm
p}$ values of these nuclei should also be little different from each
other.

    It follows from the above that for checking the hypothesis for
    the relationship between the nuclear cluster structure and the momentum
    transfer value $q_{\rm p}$, it is necessary:
    \begin{itemize}
    \item [a)] to determine the $q_{\rm p}$ value for the $^7$Li nucleus;
    \item [b)] to define more exactly the $q_{\rm p}$ value for the $^6$Li nucleus;
    \item [c)] to obtain the $q_{\rm p}$ values for the previously investigated nuclei.
    \end{itemize}

\section{ The experiment and handling of the measured data }\label{sec:1}

The measurements, from which the present \mbox{$S_{\rm L}(q)$} values were determined,
 were carried out at the experimental facility SP-95 with the use of the electron beam
 from the NSC KIPT electron linear accelerator LUE-300. The electron beam of
 monochromaticity between 0.4\% and 0.6\%, and of energies ranging from 104 to 259 MeV,
 was incident on the $^6$Li (or $^7$Li) target, the isotopic enrichment of
  which in the nuclide of interest was determined to be 90.5\% (or 93.8\%),
  respectively. The measurements were performed at electron scattering angles
  from 34.2$^{\circ}$ up to 160$^{\circ}$. For momentum analysis of scattered
  electrons we have used the spectrometer that had the second-order
  double focusing in vertical and horizontal planes \cite{9}.
  Electrons in the focal plane of the spectrometer were registered by
  the 8-channel scintillation Cherenkov counter \cite{10}.
%  The description of the facility has been given in several publications
% (e.g., see refs.~\cite{7, 8, 11, 12}).

The experimental setup has been
described a number of times in the literature, see e.g. 
\cite{7, 8, 11, 12}. A detailed description of the measurements and
the data processing is presented in refs. [3, 6, 7, 8].

The experiment was designed so that the response functions at
several constant 3-momentum transfer values ranging from
0.750 to 1.625 fm$^{-1}$, and also, the Coulomb sums corresponding
to these functions, could be obtained from the measurements. It should
 be mentioned that the most complicated and labor-consuming stage
 in these experiments is the processing  of the measurement results
 for obtaining the response functions and the Coulomb sums.
 Taking into account the long duration of the processing, the
 work was planned so as to process first the data measured at
 the highest $q$ values, and then to process the data corresponding
 to lower momentum transfers. One of the advantages of this approach
 was the point that if the processing of a part of the experimental
 data yielded the physical data of prime interest, they could be
  discussed and submitted for publication at once, without waiting
  for the final processing of the whole body of initial measured data.

At the previous stage of  measured data processing, we have obtained
in this way four \mbox{$S_{\rm L}(q)$} values for $^7$Li at
$q = 1.250 \div 1.625$ fm$^{-1}$ \cite{8}, and five \mbox{$S_{\rm L}(q)$}
values for $^6$Li at $q = 1.125 \div 1.625$ fm$^{-1}$ \cite{7}.

\begin{figure}[bh] %\noindent
\centering{
\includegraphics[width=65mm]{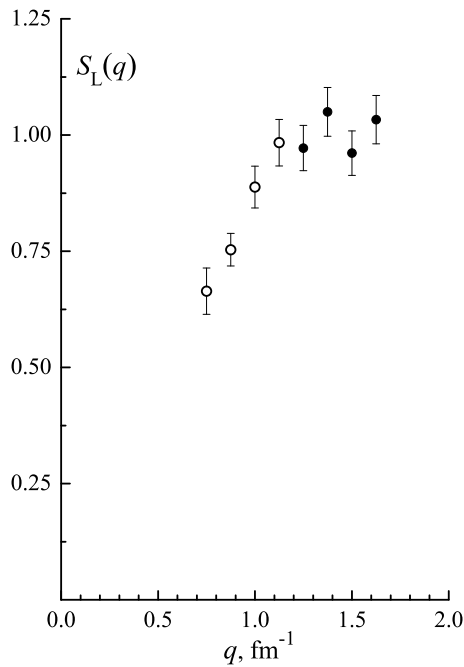}
} \caption{Coulomb sum of $^7$Li. Full squares -- $^7$Li \cite{8};
open squares -- $^7$Li (present data).} \label{fig2}\end{figure}

\begin{figure}[bth] %\noindent
\centering{
\includegraphics[width=65mm]{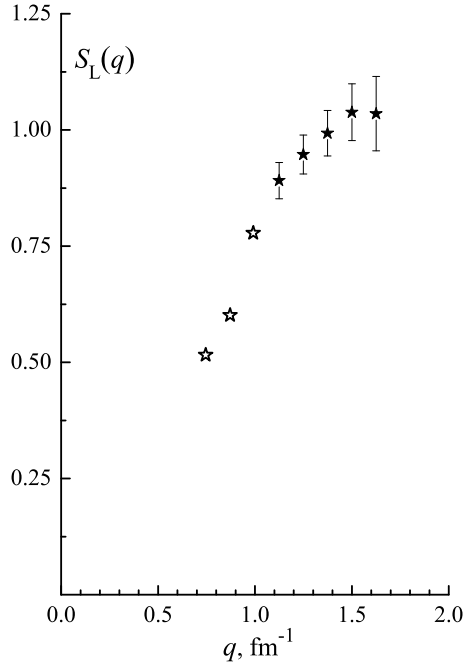}
} \caption{Coulomb sum of $^6$Li. Full circles -- $^6$Li \cite{7};
open circles -- $^6$Li (present data).} \label{fig3}\end{figure}

By the present time, in addition to the above-given values, we have
obtained \mbox{$S_{\rm L}(q)$} values for $^7$Li at $q = 0.750 \div
1.125$ fm$^{-1}$ (Fig.~\ref{fig2}), and preliminary \mbox{$S_{\rm
L}(q)$} values for $^6$Li at $q = 0.750 \div 1.000$~fm$^{-1}$
(Fig.~\ref{fig3}).

\section{ NUCLEAR CLUSTER STRUCTURE  AND THE COULOMB SUM }\label{sec:2}

\begin{figure}[bth] \centering{
\includegraphics[width=60mm]{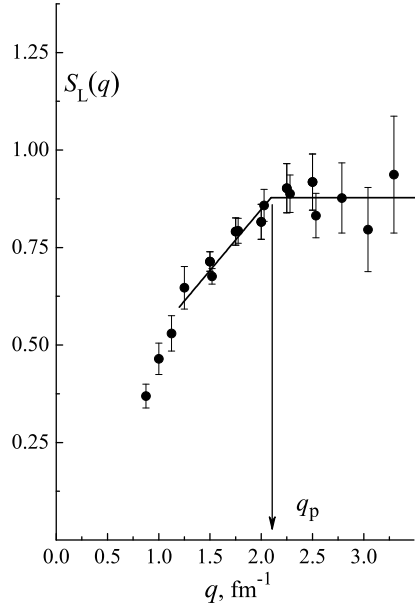}
} \caption{Coulomb sum of $^4$He. Full circles -- $^4$He
\cite{3,4,5}; horizontal and inclined lines --  data fitting; the
intersection of the lines determines the $q_{\rm p}$ value.}
\label{fig4}\end{figure}

 To analyse the relationship between the
momentum transfer $q_{\rm p}$ and the nuclear cluster structure, the
$q_{\rm p}$ value determination
 must be formalized using a certain simple procedure, which will be
 applied to the experimental \mbox{$S_{\rm L}(q)$} values of the nuclei
  under consideration. We define $q_{\rm p}$ as the momentum transfer
  that corresponds to the point of intersection of two straight lines,
  one of which (horizontal) approximates the \mbox{$S_{\rm L}(q)$}
  values on the plateau of \mbox{$S_{\rm L}(q)$},
  and the other line approximates the \mbox{$S_{\rm L}(q)$} values before
  reaching the plateau starting from \mbox{$S_{\rm L} \approx 2/3S_{\rm L,p}$},
  where \mbox{$S_{\rm L,p}$} is the \mbox{$S_{\rm L}(q)$} value on the plateau.
  The given definition of the momentum transfer $q_{\rm p}$ is exemplified
  by the \mbox{$S_{\rm L}(q)$} for the $^4$He nucleus (see Fig.~\ref{fig4}).

\begin{figure*}[th] \centering{
\includegraphics[width=130mm]{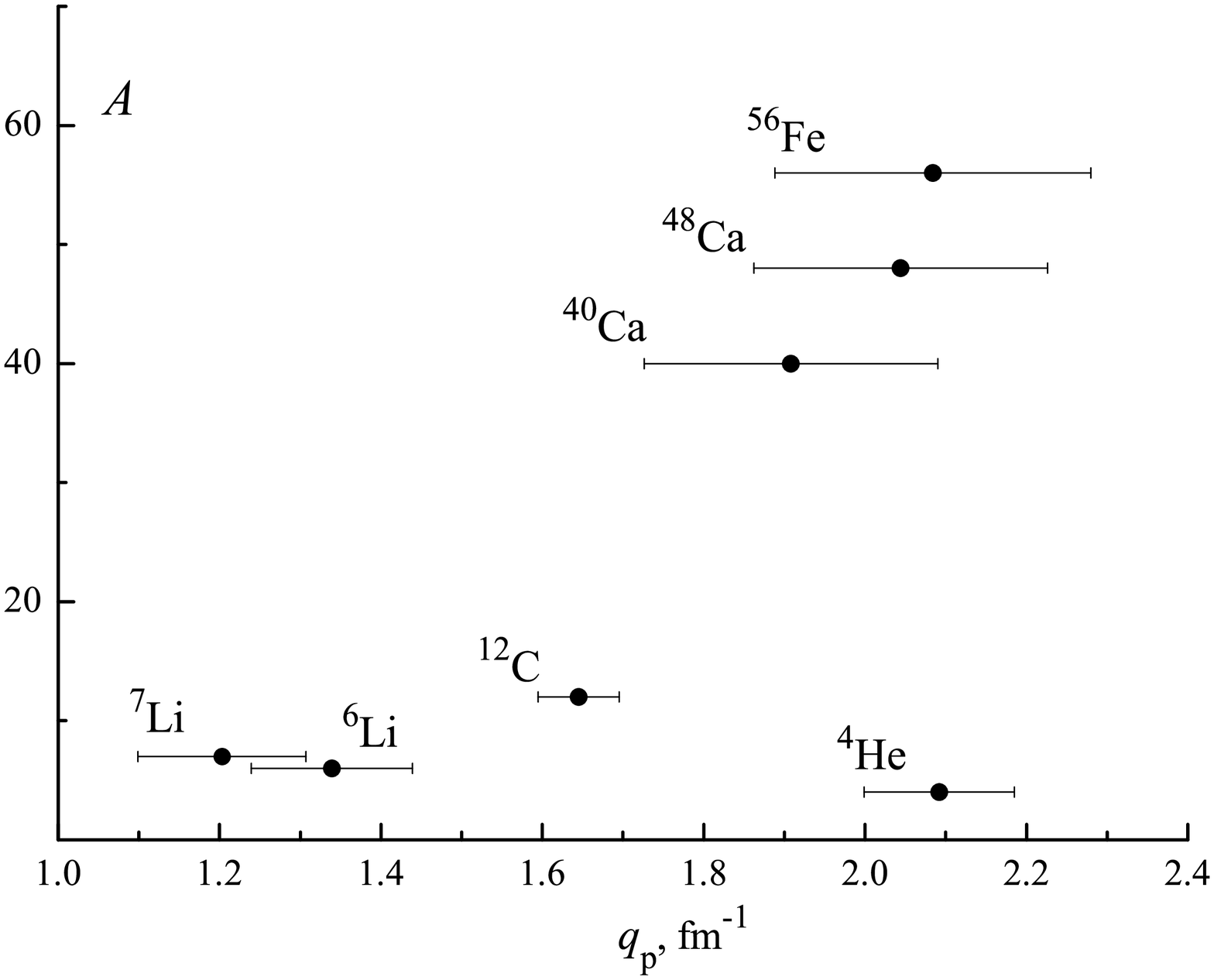}
} \caption{Momentum transfers $q_{\rm p}$ for different nuclei. The
atomic mass $A$ is plotted on the axis of ordinates.} 
\label{fig5}\end{figure*}

 We apply this definition of $q_{\rm p}$ to
all the nuclei having the atomic mass $A \geq 4$, for which a
sufficient amount of experimental \mbox{$S_{\rm L}(q)$} data is
known. These are the data of the present work and of our previous
works on the nuclei $^{6,7}$Li \cite{6,7,8}, $^4$He \cite{3} and
$^{12}$C \cite{13}. Besides, from ref.~\cite{5}, we have used the
experimental \mbox{$S_{\rm L}(q)$} data obtained at the Saclay and
Bates Laboratories for $^4$He, $^{12}$C, $^{40}$Ca, $^{48}$Ca,
$^{56}$Fe. The momentum transfers $q_{\rm p}$ derived from these
data are
 shown in Fig.~\ref{fig5}. It can be seen that the $q_{\rm p}$
 values of the nuclei $^4$He, $^{40}$Ca, $^{48}$Ca, $^{56}$Fe are
 grouped at $q_{\rm p} = (1.9 \div 2.1)$~fm$^{-1}$, and in
 the case of $^6$Li and $^7$Li - at $q_{\rm p} = (1.20 \div 1.35)$~fm$^{-1}$.
  For the $^{12}$C nucleus we have $q_{\rm p} = 1.65$ fm$^{-1}$.
   The momentum $q_{\rm p}$ grouping of the nuclei, observed in
   Fig.~\ref{fig5}, corresponds to their distribution over the
   cluster isolation parameter $x$\footnote{The parameter
    $"x" $ defines the degree, to which
   the clusters are formed within the nucleus \cite{14}.
   The $x$ value varies from $x = 1$ (shell model, e.g., $^4$He)
   to $x = 0$ (limiting case of the cluster model).}.
   The first-group nuclei are not clusterized, whereas the
   second-group nuclei are strongly clusterized. Thus, for the $^6$Li
   nucleus, the parameter $x$ varies between 0.3 and 0.4 \cite{6,14,15},
   while for $^7$Li we have $x = 0.5$ \cite{14}. With this approach, we
   arrive at understanding of the intermediate (between the two groups) value
   $q_{\rm p} = 1.65$~fm$^{-1}$ of the $^{12}$C nucleus, because this
   nucleus is though clusterized but to a less degree than the nuclei
   of the lithium isotopes.  For the $^{12}$C
   nucleus, the parameter $x$ ranges from 0.7 to 0.8 \cite{14,15}.

\begin{figure}[th] \centering{
\includegraphics[width=65mm]{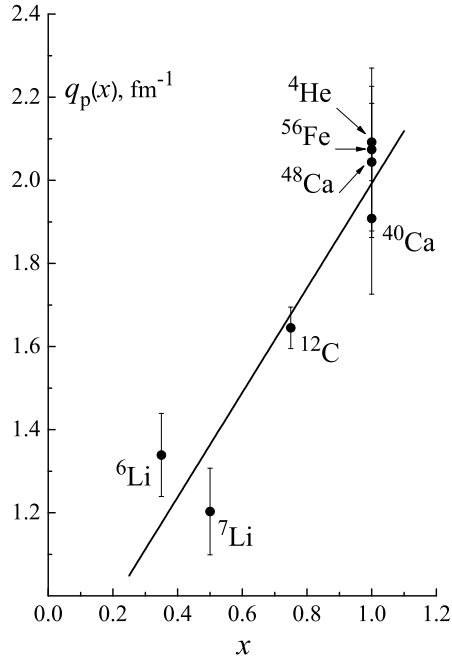}
} \caption{Momentum transfer $q_{\rm p}$ versus the cluster
isolation parameter $x$ for different nuclei. The straight line
represents the data fitting by the linear dependence.}
\label{fig6}\end{figure}

 Let us consider the momentum $q_{\rm p}$ as
a function of the parameter $x$. For this purpose we put the cluster
isolation parameter of the nuclei $^4$He, $^{40}$Ca, $^{48}$Ca,
$^{56}$Fe to be equal to 1.0. As is obvious from Fig.~\ref{fig6},
the $x$ dependence of $q_{\rm p}$ is close to linear, this being in
agreement with the result of fitting the straight line to all the
data with the least $\chi_i^2$ value. Note that the observed
dependence displays a high sensitivity of $q_{\rm p}$ to the $x$
value.
 After refinement of $q_{\rm p}(x)$\footnote{In particular, more precise determination 
 of $x$ values for the nuclei of lithium isotopes.},
 this feature of the function considered might be used for determination
 of $x$ from the $q_{\rm p}$ value. However, because of the laborious
 procedure of obtaining experimental Coulomb sums, this method would be
 hardly applicable in practice.

\section{ RESULTS AND CONCLUSIONS} \label{sec:res}

The results of the present work can be summarized as follows.
\begin{itemize}
    \item [A.] Experimental \mbox{$S_{\rm L}(q)$} values of the nuclei $^6$Li and $^7$Li have been obtained at momentum transfers $q = 0.750 \div 1.000$~fm$^{-1}$ and $q = 0.750 \div 1.125$~fm$^{-1}$, respectively. This has essentially extended the range of the measured \mbox{$S_{\rm L}(q)$} towards $q$ values lower than those investigated in refs.~\cite{7,8}.
    \item [B.] Using the \mbox{$S_{\rm L}(q)$} data of the present work and of works \cite{6,7,8}, the momentum transfer $q_{\rm p}$ has been determined for the $^7$Li nucleus ($q_{\rm p} = 1.20 \pm 0.10$~fm$^{-1}$), and has been redetermined more exactly for the $^6$Li nucleus ($q_{\rm p} = 1.35 \pm 0.10$~fm$^{-1}$). The analysis of the available literature data on the Coulomb sums for  $^4$He, $^{12}$C, $^{40}$Ca, $^{48}$Ca, $^{56}$Fe has yielded the $q_{\rm p}$ values for the mentioned nuclei (see Fig.~\ref{fig5}).
    \item [C.] The momentum transfers $q_{\rm p}$ of $^{6,7}$Li nuclei have been found to be much lower than those in the case of other nuclei.
    \end{itemize}

The comparison of the present experimental data with the data
obtained elsewhere for a number of nuclei has demonstrated the
validity of the hypothesis of the manifestation of nuclear cluster
structure  in the Coulomb sum of the nucleus.

The effect manifests itself in the observable proportionality of the
momentum transfer $q_{\rm p}$ to the cluster isolation parameter
$x$, which characterizes the degree of nuclear clusterization.
Besides, the hypothesis under discussion might be also supported
provided that the Coulomb sums were measured for the $^9$Be nucleus
at $q = 0.8 \div 1.7$~fm$^{-1}$, from which the momentum $q_{\rm p}$
of this nucleus can be derived. Since the parameter $x = 0.6$
\cite{14}, related to the $^9$Be nucleus, lies between the $x$
values of the nuclei $^{6,7}$Li and  $^{12}$C, i.e., in the range
from 1.3~fm$^{-1}$ to 1.6~fm$^{-1}$.

\end{document}